# Extraction of electrokinetically separated analytes with on-demand encapsulation

X.F. van Kooten[a,b], M. Bercovici*[b] and G.V. Kaigala*[a]

[a]IBM Research – Zurich, Rüschlikon, Switzerland
[b]Technion – Israel Institute of Technology, Haifa, Israel

## 1. ABSTRACT

Microchip electrokinetic methods are capable of increasing the sensitivity of molecular assays by enriching and purifying target analytes. However, their use is currently limited to assays that can be performed under a high external electric field, as spatial separation and focusing is lost when the electric field is removed. We present a novel method that uses two-phase encapsulation to overcome this limitation. The method uses passive filling and pinning of an oil phase in hydrophobic channels to encapsulate electrokinetically separated and focused analytes with a brief pressure pulse. The resulting encapsulated sample droplet maintains its concentration over long periods of time without requiring an electric field and can be manipulated for further analysis, either on- or off- chip. We demonstrate the method by encapsulating DNA oligonucleotides in a 240 pL aqueous segment after isotachophoresis (ITP) focusing, and show that the concentration remains at 60% of the initial value for tens of minutes, a 22-fold increase over free diffusion after 20 minutes. Furthermore, we demonstrate manipulation of a single droplet by selectively encapsulating amplicon after ITP purification from a polymerase chain reaction (PCR) mix, and performing parallel off-chip detection reactions using the droplet. We provide geometrical design guidelines for devices implementing the encapsulation method, and show how the method can be scaled to multiple analyte zones.

## 2. INTRODUCTION

Electrophoretic techniques such as isotachophoresis (ITP), isoelectric focusing (IEF) and temperature gradient focusing (TGF) are powerful methods to separate and focus charged analytes. Their versatility as both a sample preparation and preconcentration method makes them an important component of numerous bioanalytical workflows. For instance, ITP has been used as a purification method[1,2] or as an upstream preconcentration step to enhance sample loading in capillary zone electrophoresis[3,4]. Isoelectric focusing has been used as a preparative method in two-dimensional electrokinetic separations[5,6] and is also an effective tool for separation and mapping of proteins[7].

In conventional techniques, such as slab-gel electrophoresis, it is relatively easy to access the separated or focused sample for further analysis, *e.g.* via band excision or blotting. Although the migration of electrophoretic techniques to the microchannels of lab-on-a-chip devices has provided significant advantages such as smaller reagent volumes, less Joule heating and better control over sample injection, the format of closed channels complicates access to separated analytes for downstream analysis. A notable exception is the use of ITP as a purification method[2,8], in which the purified analytes run through the full length of the channel and can be extracted from the LE reservoir, albeit at the cost of sample dilution during manual handling.

At the same time, in-channel detection is limited primarily to optical methods, as the high electric fields used in separations preclude the use of exposed electrical sensors (*e.g.* galvanic conductivity[9,10] or surface plasmon resonance[11] sensors). Additionally, the chemical conditions required for the two steps (separation and detection) may differ significantly[12].

Enabling compatibility of on-chip electrophoretic methods with downstream processing and detection requires a method that maintains spatial concentration profiles even after the electric field is removed. Zare *et al.* achieved this by trapping focused analytes using elastomeric valves[13,14]. While this approach does maintain spatial separation and focusing of analytes, it does not allow extraction of analytes from the main channel without dispersion. An alternative approach, first shown by Chiu *et al.*[15] and later, using pressure-driven flow, by Edel *et al.*[16], involves generating a continuous stream of aqueous droplets from the separation phase during electrophoresis. While this approach is elegant because the indiscriminate droplet generation requires no feedback control, its main drawback is the large number of 'empty' droplets (not containing analytes) and the challenge to track and manipulate desired droplets. Furthermore, the indiscriminate generation of droplets may lead to analyte zones being split between multiple droplets. This makes the continuous droplet approach unsuitable for assays that require pure and highly concentrated analyte fractions. Ismagilov *et al.*[17] performed IEF through a large number of serially connected chambers in a SlipChip, and then segmented the separation phase in a single step by breaking the connection between chambers, but this approach does not overcome the issue of indiscriminate splitting, and does not allow individual droplets to be manipulated.

We here present an alternative method, in which we capture and extract electrophoretically separated or focused analytes in single encapsulations created on demand. The encapsulation method is based on a two-phase interface which ensures retention of the analyte zones for tens of minutes. The resulting aqueous segment can then be delivered to downstream processes on- or off-chip, and its surface contact angle can be tuned to match the application. We here focus on the encapsulation of analytes separated and focused using ITP and demonstrate selective encapsulation of DNA amplicon from a polymerase chain reaction (PCR) mix. However, we believe that the concepts and mechanism presented here will useful for a variety of other electrophoretic techniques.

3. **OPERATING PRINCIPLE AND DESIGN CONSIDERATIONS**
*3.1. Operating principle of the on-demand encapsulation method*
Figure 1a presents the operating principle of a device implementing on-demand encapsulation. A primary channel is initially filled with an aqueous separation matrix. This separation channel is joined by two secondary side channels, into which an immiscible oil phase is introduced. The oil fills the channels by capillary action (Fig. 1b) but stops at capillary valves located where the secondary channels meet the primary channel (Fig. 1c).

The oil interface remains pinned in this position while an electric field is applied along the primary channel to initiate electrokinetic separation and focusing (Fig. 1d). Once the desired analyte band reaches the encapsulation region, the electric field is removed, and oil is injected from the side channels, thereby 'encapsulating' the analyte in a closed segment of the aqueous separation phase (Fig. 1e). The operating principle of the method for on-demand encapsulation makes it inherently compatible with various on-chip electrokinetic techniques that use aqueous separation phases, such as zone electrophoresis, isotachophoresis and isoelectric focusing.

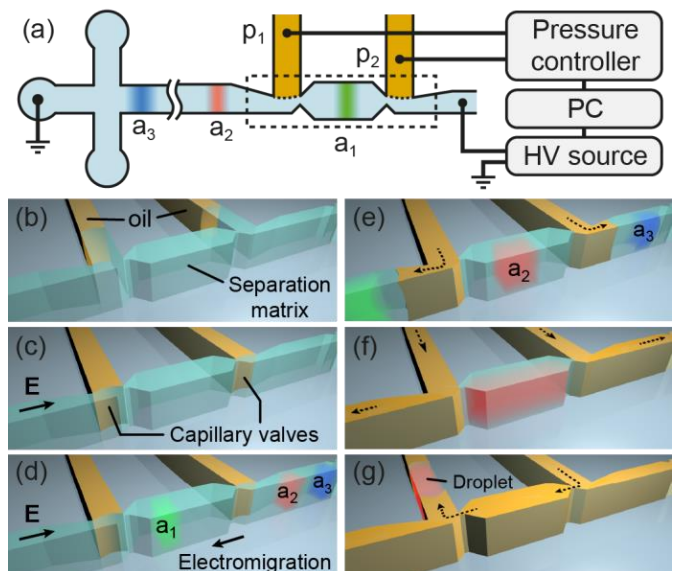

**Figure 1. Operating principle of on-demand encapsulation and extraction of electrokinetically focused analytes.** (a) Schematic of the on-demand encapsulation method, which can be implemented by adding an encapsulation geometry to a separation channel. The dashed box indicates the region that is enlarged in the following subfigures. (b) We initially fill the encapsulation device with an aqueous separation phase. We then introduce an oil phase into two side channels. The oil passively fills the side channels by capillary action, displacing the aqueous phase. (c) Capillary valves hold the oil interface in place and prevent it from entering the main channel. (d) We apply a voltage across the primary channel to electrokinetically separate and focus analytes. (e) Once the desired analyte band reaches the encapsulation region, we remove the electric field and apply a pressure pulse to the side channels, injecting the oil into the primary channel to encapsulate the analyte band as a water-in-oil segment. (f) The concentration of surfactant can be changed *in-situ* to adjust the contact angle, depending on the desired downstream application. (g) The analyte droplet is ejected from the encapsulation region by applying a differential pressure between the two side channels and can be collected for further on- or off-chip analysis.

Following encapsulation, the oil phase can be exchanged to increase or reduce the interfacial tension (Fig. 1f). This allows the contact angle of the analyte segment to be adjusted to the requirements of downstream applications. For instance, replacing the encapsulation oil with a high-surfactant oil results in an aqueous droplet with a high surface contact angle. This droplet can then be ejected from the encapsulation region and can be collected for on- or off-chip downstream analysis (Fig. 1g).

The device shown in Figure 1 has only two side channels, enabling encapsulation of a single pure analyte band or a mixture of several analyte bands. In this work, we use this device to demonstrate the concept of on-demand encapsulation. However, the encapsulation-on-demand method is not limited to a single encapsulation region as each oil injection point is independent of its neighbor. As such, the unit shown in Figure 1 can be extended to encapsulate *n* analyte zones using *n+1* side channels. As shown in SI Figure S1, when encapsulating more than one analyte zone, a 'cross-over' geometry is preferred, in which each central side channel has a counterpart opposite it in the main channel. Such a geometry minimizes displacement of the aqueous phase during oil injection, enabling encapsulation of multiple sharp analyte zones. The extension to multiple analyte zones is limited only by the number of pressure lines and valves, which increases linearly with the number of encapsulated analytes.

### 3.2. Geometrical design considerations for devices implementing on-demand encapsulation

The geometry for on-demand encapsulation shown in Figure 2 is based on interfacial pinning in two stable positions at the end of the side channels. The pinning position (A-B in Figure 2a) holds the oil interface during separation, whereas the second pinning position (B-C in Figure 2a) ensures that the oil does not penetrate the volume containing the analyte during the encapsulation step. This minimizes the displacement of the analyte zone during encapsulation and ensures that its spatial concentration profile is maintained.

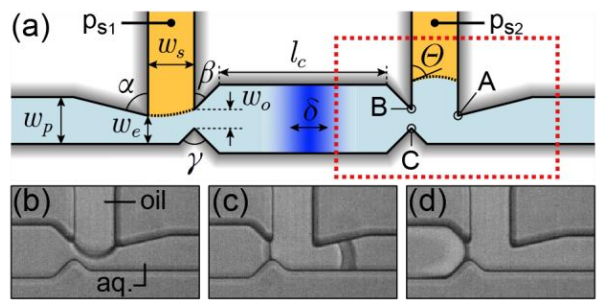

**Figure 2. Top view of the microchannel geometry showing the key design parameters for oil pinning structures enabling on-demand encapsulation.** (a) Schematic drawing of angles and dimensions for contact line pinning of the oil-water interface. The interface first pins between points $A$ and $B$, and upon application of a pressure burst exceeding the Laplace pressure, pivots to pin between points $B$ and $C$. The angles $\alpha$, $\beta$ and $\gamma$ relate to static pinning behavior. The dimensions $w_e$, $w_s$ and $w_s$ determine the behavior of the oil interface when pressure is applied. In this device, $\alpha = 77°$, $\beta = 45°$, $\gamma = 90°$, $w_s = 30\ \mu m$, $w_e = 15\ \mu m$ and $w_o = 7.5\ \mu m$. (b) Microscope image of the encapsulation region indicated by the dashed red box in Figure 2a, showing expansion of the oil interface into the aqueous (aq.) phase at the right capillary valve as pressure is applied. (c) The interface pins in a new stable position after limited displacement of the aqueous phase in the direction of the encapsulation region. (d) Continued injection of oil into the primary channel enables the contact angle to be adjusted by varying the surfactant concentration in the oil phase. An oil film can be seen surrounding the analyte droplet in the encapsulation region.

*Pinning at the first capillary valve*

The first set of capillary valves (A-B) is defined by angles $\alpha$ and $\beta$, as indicated in Figure 2a. On hydrophobic (oleophilic) surfaces, the contact angle $\theta$ is less than 90°, and the contact line will remain pinned if $\alpha, \beta < 90°$. However, we recommend a safety margin for two reasons. First, the corners of structures fabricated by soft lithography are often slightly rounded, an effect that can significantly alter the behavior of pinned interfaces[18]. Second, the capillary pressure of the unconstrained out-of-plane contact line will change the effective in-plane contact angle during pinning. We observed that this causes the oil phase to creep along the corners of the channel, leading to spontaneous bursting of the capillary valve for angles of $\alpha$ and $\beta$ close to 90°. The lower bound of $\alpha$ and $\beta$ is given by fabrication, with 45° being the practical limitation for PDMS structures. Figure 2b shows the oil interface at the first capillary valve, with $\alpha = 77°$ and $\beta = 45°$, which remains stable for tens of minutes.

*Pinning at the second capillary valve*

Once the oil is pushed into the separation channel by a brief pressure pulse (Fig. 2c), the interface progresses to a second stable position at a capillary valve defined by points $B$ and $C$. This re-pinning is important as it loosens the timing constraints for synchronization of the pressure pulses being applied to the side channels. By designing $\alpha > \beta$, the interface first de-pins from $A$ and re-pins at point $C$ while 'pivoting' and remaining pinned at point $B$ (Fig. 2d). Pivoting of the oil interface around point $B$ serves to minimize analyte displacement by preventing any lateral movement of the oil in the direction of the encapsulation region. The design constraints for correct operation are therefore $45° \leq \beta < \alpha \leq 90°$, which is also satisfied by choosing $\alpha = 77°$ and $\beta = 45°$. Supplementary Figure S4 shows no displacement of the analyte when one capillary valve bursts before the other.

### *Dimensions for controlling burst pressure*

The 'burst pressure' of a capillary valve is equal to the Laplace pressure of the interface, given by the Laplace-Young equation, $\Delta p_L = \Gamma(\frac{1}{R_{xy}} + \frac{1}{R_z})$, where $\Gamma$ is the interfacial tension, and $R_{xy}$ and $R_z$ are the in-plane and out-of-plane curvature radii of the interface, respectively. As the channels have a uniform depth of 50 $\mu m$, the key geometrical parameters determining the curvatures are the width of the side channel ($w_s = 30\ \mu m$), the extraction gap ($w_e = 15\ \mu m$) and the encapsulation orifice (designed as 5 $\mu m$, fabricated as $w_o = 7.5\ \mu m$). Choosing $w_o < w_s$ ensures that $\Delta p_{L,\widehat{AB}} < \Delta p_{L,\widehat{BC}}$, so that the interface re-pins at the second capillary valve if the applied pressure does not exceed the Laplace pressure of the capillary stop between $B$ and $C$, $\Delta p_{L,\widehat{BC}}$. Furthermore, choosing $w_o < w_e$ ensures that the capillary pressure at the second pinning location is greater than at any other location in the separation channel, and as a result any injected oil is preferentially directed away from the encapsulation zone rather than towards it.

Following encapsulation, the analyte zone can be ejected from the encapsulation region into one of the side channels. The minimal differential pressure required to push the encapsulated segment through the orifice between $B$ and $C$ is equal to the Laplace pressure of the pinned interface, *i.e.* $p_{s1} - p_{s2} \geq \Gamma(\frac{2}{w_o} + \frac{2}{h_c})$. Following ejection, the encapsulated segment will be pushed upward into the side channel if $w_e < w_s$. If $w_e = w_s$, splitting may occur, with the larger portion of the segment being pushed into the branch with the lower hydraulic resistance.

### *Minimum dimensions of the encapsulation region*

The volume of the encapsulation chamber should be as small as possible to retain the highest concentration of analyte. As such, the length of the encapsulation chamber should be similar to the length of the analyte zone, $l_c \approx \delta$. The lower limit on $l_c$ is determined solely by the response time of the voltage or pressure control system, which should be fast enough to execute the encapsulation while the analyte zone is within the chamber. The geometry shown in Figure 2a allows a minimal encapsulated volume of 50 $pL$ (for $l_c = 0$), although this volume may be further reduced by using shallower channels.

## 4. MATERIALS AND METHODS
### *4.1. Device fabrication*

The implementation of on-demand encapsulation using an oil/water system requires an oleophilic surface of the channel. Additionally, the contact angle must be greater than zero to enable contact line pinning at the capillary valves. Therefore, our choice for Fluorinert FC-40 as an inert oil phase excluded the use of glass as a substrate material, as $\theta \sim 0°$[19]. Instead, we fabricated the devices on 250 $\mu m$ thick transparent silicone rubber (HT-6240, MVQ Silicones GmbH, Weinheim, Germany), which we bonded to a glass microscope slide after 30 $s$ air plasma treatment using a handheld corona pen (Relyon Plasma, Regensburg, Germany)

We used a soft lithography process[20] to fabricate the channels. We made molds using SU-8 rather than using deep reactive ion-etched (DRIE) Si because the latter caused unstable pinning at the capillary valves due to partially isotropic etching of the horizontal Si surface next to vertical walls. We fabricated a negative mold by patterning 50 $\mu m$ thick SU-8 3050 photoresist on a silicon wafer. We then cast a 4 $mm$ thick layer of polydimethylsiloxane (PDMS) onto the wafer in a 10:1 ratio of monomer base to crosslinking agent, degassed it under vacuum for 20 $min$ and cured it for 2 $h$ at 80 °C. We then released the PDMS from the wafer and punched reservoirs 3 $mm$ in diameter using a disposable biopsy punch. After rinsing the surface with ethanol and drying with nitrogen, we treated the PDMS and the silicone rubber substrate with air plasma for 30 $s$ and bonded the two together. Finally, we filled the channels and reservoirs with FC-40 oil containing 1% v/v trichloro(octyl)silane (TCS) and baked the devices for a further 2 $h$ at 120 °C.

We obtained SU-8 3050 from Microchem (Westborough, MA), Si wafers from Si-Mat (Kaufering, Germany) and PDMS from Dow Corning (Midland, MI). We purchased FC-40, TCS, PFO, bistris, HCl, tricine, imidazole, PVP and MOPS from Sigma-Aldrich (Buchs, Switzerland), SYBR Green I from Thermo Fisher Scientific (Waltham, MA) and Picosurf-1 surfactant from Dolomite Microfluidics (Royston, United Kingdom). We obtained RotorGene SYBR Green master mix and RNase free water from Qiagen (Hombrechtikon, Switzerland), and all DNA oligonucleotides (sequences provided in SI Section S5) from Integrated DNA Technologies (Skokie, IL).

### 4.2. Device operation

We first rinsed the channels with ethanol and deionized water for 1 min each, and then flowed leading electrolyte (LE) for 15 min by filling all reservoirs with LE and applying a vacuum from the west reservoir. For ssDNA focusing, the LE consisted of 200 mM bistris, 100 mM HCl and 1% m/v polyvinylpyrrolidone (PVP). For separation of polymerase chain reaction (PCR) products, the LE contained 200 mM imidazole, 100 mM HCl, 7% PVP and 1× SYBR Green I.

We removed the LE from the north reservoirs and replaced it with the oil phase, consisting of Fluorinert FC-40 and 2% v/v 1H,1H,2H,2H-perfluoro-1-octanol (PFO). The oil passively filled the side channels with a velocity of ~15 $\mu m/s$. Once the interface had reached approximately 1 mm into the channel, we emptied the north reservoirs and refilled them with FC-40 containing 0.5% Picosurf-1 surfactant. Depending on the desired contact angle of the final analyte segment a different surfactant or surfactant concentration may be used, or this step may be skipped entirely.

After replacing the oil, we connected pressure control lines to the reservoirs. These lines were connected to a two-way valve (ESS 2-SWITCH, Fluigent, France), of which one inlet was at atmospheric pressure and the other inlet was connected to a controlled pressure source (MFCS-EZ, Fluigent, France). Supplementary Figure S3 shows the experimental setup.

Once the oil interface in both side channels reached its stable pinned position at the capillary valve, we removed the LE in the west reservoir, rinsed the reservoir with deionized water and filled it with the terminating electrolyte (TE) containing the sample to be focused or separated. For ITP focusing of ssDNA, we spiked fluorescently labeled oligonucleotides (5'-/5Alexa-546N/GA TGG GCC TCC GGT TCA TGC CGC CCA -3') at 5 nM in a TE consisting of 20 mM bistris, 10 mM tricine, 1% m/v PVP. For post-PCR separation we diluted the PCR product 1:1 in TE, for a final TE concentration of 40 mM imidazole, 20 mM tricine, 1% m/v PVP and 2 mM 3-(N-Morpholino)propanesulfonic acid (MOPS).

We then applied a voltage between the cathode in the TE and the anode in the LE using a high-voltage power supply (2410, Keithley, Cleveland, OH). We used a custom MATLAB script to simultaneously control the voltage, measure the current, set the pressure and control the two-way valves. For ssDNA focusing, the electric field was 300 V/cm. For post-PCR separation, we applied 40 V/cm for 60 s, replaced the sample in the west reservoir with pure TE without MOPS, and continued to apply 80 V/cm.

We optically tracked the electromigrating analyte bands, and manually triggered the script to start the encapsulation sequence once the desired ITP interface reached the encapsulation region. This sequentially set the voltage to zero and switched the valves on and off, so that the burst pressure was briefly applied to the side channels. For the encapsulation of ITP-focused ssDNA we applied 26 mbar for ~80 ms, and for encapsulation of separated PCR products we applied 40 mbar for 1.5 s to displace the viscous separation buffer.

### *4.3. PCR and off-chip amplification assay*

We performed two rounds of PCR: the first to amplify for the $\beta$-actin sequence in genomic DNA, and the second after ITP separation and performing on-demand encapsulation, as an assay for the presence of $\beta$-actin amplicon in the encapsulated analyte band. The cell culture, lysis and purification protocols to obtain gDNA are provided in SI Section S4. Details of the PCR reactions are provided in SI Section S5.

Following ITP separation and encapsulation, we applied 10 mbar to both side channels for 10 min to continuously flow FC-40 containing Picosurf-1 surfactant. We then ejected the encapsulated analyte by applying a differential pressure of 10 mbar between the left and right side channel. Finally, we manually collected the ejected droplet from the right reservoir in 3 $\mu L$ of oil using a pipette, transferred it to a 200 $\mu L$ vial, and stored it at $-20°C$ for further processing.

### 5. RESULTS AND DISCUSSION
### *5.1. Retention of focused analyte using on-demand encapsulation*

We characterized the ability to retain highly concentrated analytes after encapsulation. Figure 3a and b respectively show the filling and passive pinning of an oil phase in the side channels. We performed ITP focusing of fluorescently labeled single-stranded DNA oligonucleotides ($L = 26\ nt$) in the primary channel (Fig. 3c). The ssDNA focused into a 113 $\mu m$ wide moving zone (measured at 10% of the maximum concentration). As this band passed through the encapsulation region we manually triggered the injection of oil to encapsulate the analyte band in an aqueous segment with a total volume of 240 pL (Fig. 3d). Supplementary video S1 shows ITP focusing, encapsulation and ejection of oligonucleotides in an encapsulation device.

We estimated the pressure required to inject the oil to be 60 mbar using the Laplace-Young equation. However, we expected this value to be an overestimation, as the interfacial tension for water in FC-40 reported by Mazutis and Griffiths[21], $\Gamma \approx 52\ mN/m^2$, does not account for the presence of surfactant, and we also observed a lower oil contact angle on PDMS than on silicon rubber. We empirically determined that a pressure of just 26 mbar applied for approximately 80 ms was sufficient to reliably burst the capillary valves. We saw no change in the surface contact angle of the oil or in the required burst pressure upon applying an electric field, indicating that the higher local electric field strength in the constrictions near the oil interface did not substantially affect the interfacial tension. Furthermore, although we did not observe any irregularities in the ITP focusing using our experimental conditions (see Supplementary video S1), temperature-induced mobility changes may occur if high electric field strengths are applied.

Following encapsulation, we tracked the fluorescence in the ssDNA segment. Fig. 3f shows the concentration profile along the primary channel 0, 1.5 and 15 min after encapsulation, normalized to $C_0$, the intensity at $(x, t) = (0,0)$. An initial Gaussian concentration profile rapidly diffused to a uniform concentration within the encapsulated segment, so that after 1.5 min the maximum concentration is 54% of the initial value, and remained approximately constant for tens of minutes.

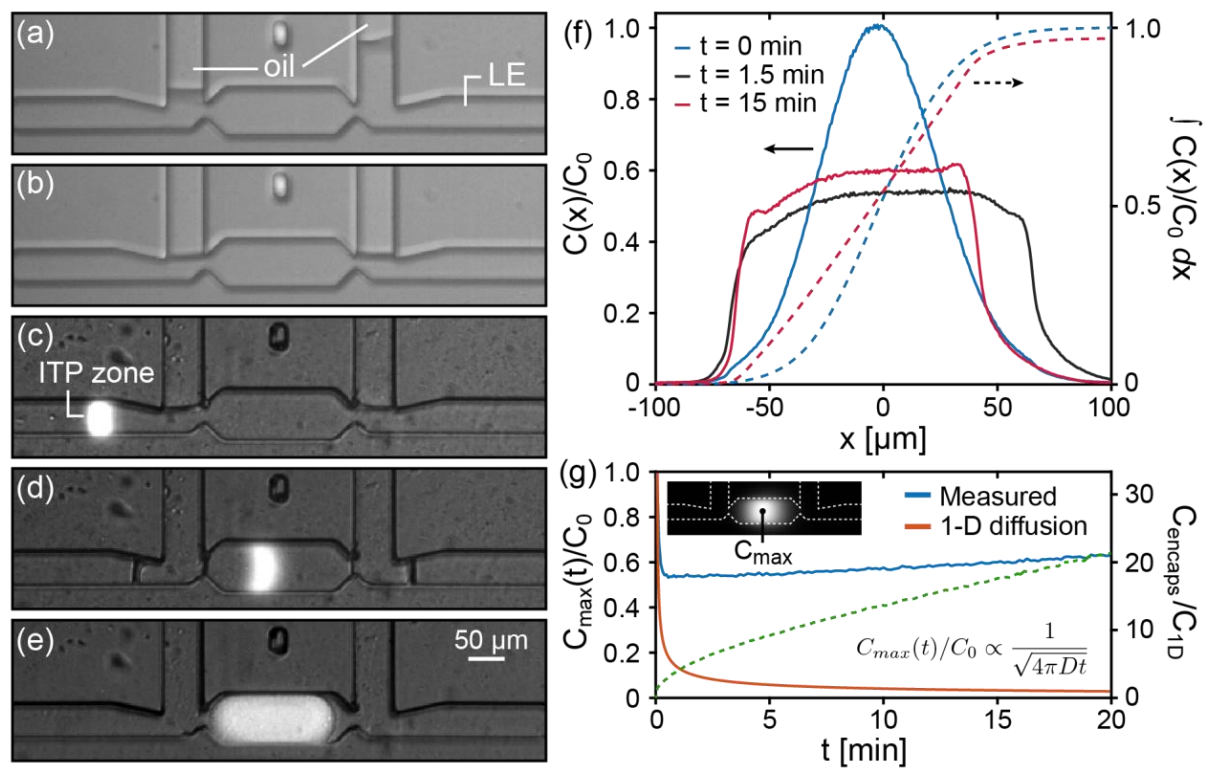

**Figure 3. Experimental results of encapsulation and extraction of ITP-focused ssDNA.** The encapsulation device is initially filled with LE (200 mM bistris, 100 mM HCl, 1% m/v PVP). (a) We replace the LE in the side reservoirs with a fluorinated oil phase (Fluorinert FC-40 + 2% v/v perfluorooctanol), which passively fills the side channels. (b) The oil interface pins at the capillary stops and remains stable for tens of minutes. (c) We then replace the LE in the west reservoir with TE (20 mM bistris, 10 mM tricine) containing 5 nM 26-nt ssDNA labeled with Alexa 546, and apply an electric field of 300 V/cm. The ssDNA focuses at the moving interface between the TE and LE. (d) Once the focused ssDNA zone reaches the encapsulation region, we remove the electric field and apply a pressure pulse (26 mbar, 80 ms) to both side channels. This causes the oil interface to enter the primary channel, encapsulating the ssDNA band in a droplet with a volume of 240 pL. The oil/water interface now reaches two new stable positions on either side of the analyte droplet. (e) We then flow a second oil phase (Fluorinert FC-40 + 0.5% Picosurf-1 surfactant) through the side channels to further increase the contact angle of the droplet and release it from the surface. (f) The concentration profile along the primary separation channel after encapsulation is initially Gaussian, but rapidly diffuses to a uniform concentration profile within the droplet. The concentration increases over time due to shrinkage of the droplet as a result of diffusion of water into the oil and PDMS. The normalized integral of the concentration (right axis) shows a slight decrease in total signal after 15 min, which may be caused by a combination of adsorption of analyte to the PDMS, partitioning into the oil phase or photobleaching. (g) Maximum fluorescence intensity of ITP-focused ssDNA following encapsulation, compared to the analytical prediction of one-dimensional diffusion without encapsulation. Images were taken at 10 s intervals, with no illumination between sampling points.

Upon injection of a second oil phase (FC-40 + 0.5% Picosurf-1) after encapsulation, the contact angle of the encapsulated segment increased, forming a droplet (Fig. 3e). Following this step, we observe a gradual decrease in the volume of the droplet. Droplet shrinkage in PDMS microchannels is reportedly caused by absorption of the aqueous phase by PDMS[22] and dissolution into the oil phase[23,24]. Although using a higher concentration of surfactant reduces the aqueous/PDMS contact area by increasing the contact angle, water uptake by the oil phase is enhanced at surfactant concentrations above the critical micelle concentration (CMC) and under continuous flow[23,24]. As shown in Figure 3f This shrinkage of the droplet also causes a slight increase of the analyte concentration within the droplet, confirming other findings that use such shrinkage as a means of analyte enrichment[25].

We note that the injection of a different oil is optional and can be varied depending on the application. Although a surfactant-stabilized analyte droplet with a high contact angle is desired for most downstream steps, certain analyses may require the analyte to be in contact with the surface. Examples include bead-based surface immunoassays or electrochemical sensors.

Figure 3f also shows the cumulative integral of the concentration along the primary channel, $\int_{x=-100}^{x'} C(x)/C_0 \, dx'$, as a measure of the total amount of analyte between $x = -100 \, \mu m$ and $x = x'$. We observed a total analyte loss of just 3% after $t = 15 \, min$, which may be due to adsorption of DNA to the PDMS or to the oil interface (despite the presence of a biocompatible surfactant) or photobleaching.

Figure 3g shows a comparison between the measured maximum concentration in the encapsulated ssDNA segment and the calculated maximum concentration as expected from one-dimensional diffusion in a microchannel. The concentration of a solute pulse injected at $x = x_0$ is given by[26] $c(x,t) = \frac{1}{\sqrt{4\pi Dt}} \exp\left(-\frac{(x-x_0)^2}{4Dt}\right)$. The maximum value at the center of the pulse ($x = 0$) decays as $c(t) = C_0/\sqrt{4\pi Dt}$, where $D$ is the diffusion coefficient ($1.52 \times 10^{-10} m^2 s^{-1}$ for 20-nt DNA oligonucleotides[27]). The improvement in concentration retention with on-demand encapsulation compared to 1-D diffusion is 22-fold after 20 min, as shown on the right axis of Figure 3g. This improvement can potentially enhance the performance of existing assays by increasing the reaction time at high concentrations of focused analytes. The 22-fold enhancement may be further increased by reducing the volume of the encapsulated segment as previously described.

### 5.2. Application of on-demand encapsulation to separated PCR products for off-chip analysis

Apart from trapping focused analytes to retain high concentrations, the strength of on-demand encapsulation lies in the ability to manipulate individual analyte segments without loss of concentration or spatial separation. This offers a way to bridge the gap between electrokinetic separation on the one hand, and sensitive on- and off-chip analysis on the other.

One example of downstream analysis is next-generation sequencing (NGS), a family of methods that depend on high-resolution 'libraries' of short DNA fragments to reconstruct genomic sequences. Library preparation requires a pure sample, containing only fragments of a certain range of lengths. This is commonly achieved by first amplifying the whole genome, using *e.g.* multiple-displacement amplification (MDA) or random-priming PCR, and subsequently purifying the amplification product to exclude nucleic acid fragments above a certain length. Purification typically involves size-based electrophoretic separation in an agarose gel matrix, followed by elution of the amplicon from an excised gel segment. While this step is well established for large samples, it is challenging for small concentrations and quantities of DNA, such as those encountered during single-cell analysis. Moreover, gel excision is not possible if the separation is performed on-chip.

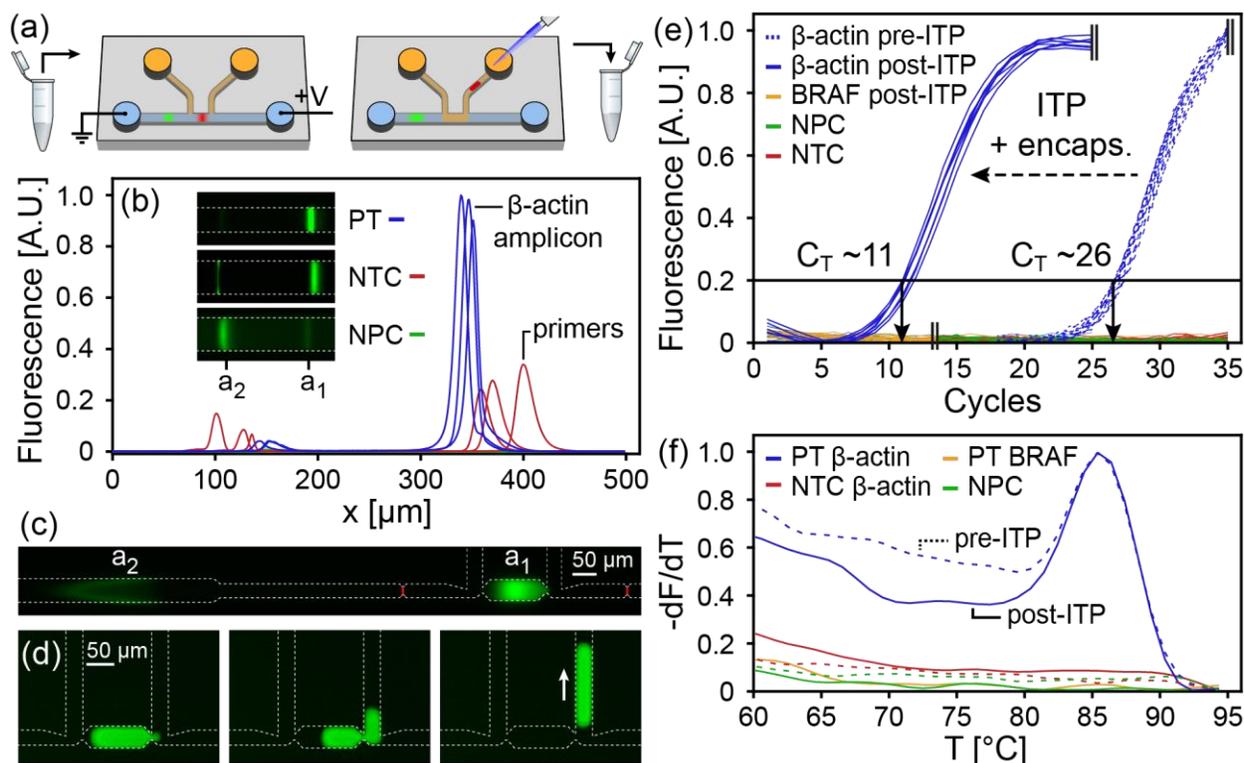

**Figure 4. Experimental results showing on-demand encapsulation of separated PCR amplicon and its extraction for use in an off-chip assay.** (a) Schematic workflow for encapsulation followed by off-chip extraction. We first perform PCR amplification of a genomic DNA template with $\beta$-actin primers. We then use ITP with an ionic spacer (MOPS) and a sieving LE (7% m/v PVP) to perform binary length-based separation of PCR products. Following separation, we encapsulate the focused and separated leading band, which contains short DNA fragments, in a 240 pL droplet. We then manually extract the droplet off-chip and perform a second, multiplexed, PCR amplification to detect the presence of amplicon in the leading band. (b) Electropherogram showing separation of amplification products from negative and positive PCR samples using ITP. Positive template (PT) samples show two peaks, whereas negative primer (NPC) and negative template control (NTC) respectively show signal only in the trailing and leading band. This indicates that the trailing band contains only genomic DNA, whereas the leading band contains primers. In the PT samples, the amplicon focuses primarily in the leading band. The inset shows fluorescence images of a PT, NTC and NPC sample in 50 μm wide channels. (c) Encapsulation of the leading analyte band containing short DNA fragments and amplicon. The encapsulated volume is 240 pL. (d) After encapsulation, we apply a differential pressure to the side channels to eject the droplet from the encapsulation zone. The droplet continues to a reservoir, from which we extract it in 3 μL of oil, and transfer it for off-chip reactions. (e) After extraction, we ran 9 independent and parallel off-chip PCR reactions (triplicates of samples containing either $\beta$-actin primers, BRAF15 primers or no primers) on each extracted droplet to detect the presence of the amplicon. $\beta$-actin amplicon is detected after 11 cycles after separation and focusing, compared to 26 cycles for the initial amplification of genomic DNA. By contrast, samples containing BRAF15 primers do not amplify even after 35 cycles (data truncated). The samples without added primers also did not amplify, indicating that no primers are carried over from the first round of PCR. (f) The derivative of the melt curve shows a peak at ~85.5 °C for positive template (PT) samples after both the first (pre-ITP) and second (post-ITP) round of PCR amplification indicating that the amplification is specific to $\beta$-actin. No peak is visible at 79 °C, the melting temperature of BRAF15 amplicon. The derivatives within each family (pre-, post-ITP, respectively shown as dashed and solid

lines) are normalized to the maximum of the positive sample in that family. The curves represent the mean of $n = 4$ curves for pre-ITP (first round) and $n = 9$ curves for post-ITP (second round) amplification.

We demonstrate how on-demand encapsulation coupled with ITP-based separation can be used to extract a highly concentrated segment of purified short DNA fragments for delivery to an off-chip amplification assay. We use ITP as it has previously been used as an effective purification technique[2,28] for sample volumes on the order of $10~\mu L$, while also providing sharp, focused analyte bands with a length on the order of $\sim 100~\mu m$. Figure 4 schematically shows the workflow. We first lyse MCF-7 cells and column-purify the lysate to extract genomic DNA. After purification we use PCR to amplify a sequence coding for $\beta$-actin, and electrokinetically separate the $\beta$-actin amplicon ($L = 202~bp$) from genomic DNA in the post-amplification mix on an encapsulation chip.

ITP-based separation can be achieved by choosing a TE with a higher mobility than the undesired species in the sample, or by adding an ionic spacer species that will focus between analyte fractions of different mobilities. We used Peakmaster[29] to estimate the mobility of MOPS, HEPES, MES and isocaproic acid spacers in terminating electrolytes containing one of several trailing ions (HEPES or tricine) and counterions (bistris, tris or imidazole). Of these candidates, we empirically determined MOPS to be a suitable ionic spacer to separate the amplicon from proteins and non-target genomic DNA.

To increase the mobility difference between short and long DNA fragments, we used a sieving LE containing 5-7% PVP. PVP has previously been used as a low-viscosity sieving matrix for post-PCR separation of DNA fragments[30] and also serves to reduce electroosmotic flow in isotachophoresis[31]. The electrophoretic mobilities of the ionic species calculated using Peakmaster were $9.1 \times 10^{-9} m^2 V^{-1} s^{-1}$ for the terminating ion (tricine), $19.7 \times 10^{-9} m^2 V^{-1} s^{-1}$ for the spacer (MOPS) and $68.5 \times 10^{-9} m^2 V^{-1} s^{-1}$ for the leading ion (Cl$^-$).

The PCR yielded amplification of $\beta$-actin from an initial gDNA concentration of $350~pg/\mu L$ after approximately 26 cycles. Figure 4b shows the electropherogram of ITP separation and focusing of PCR samples, with fluorescence detection of SYBR Green I intercalation. The addition of a MOPS spacer shows clear separation of the positive PCR samples into a trailing and a leading band. This contrasts with the negative primer (NPC) and negative template (NTC) controls, which did not amplify and respectively focus exclusively in either the trailing or leading band (Fig. 4b, inset). This shows that the trailing band in the NPC samples contains only longer genomic DNA, whereas the leading band in the NTC samples contains only primers. The signal of NPC samples in ITP was negligible compared to that of NTC samples. Given that no dsDNA is present in the NTC samples, and the melt analysis shows no primer dimer formation, we attribute the signal of the leading band in NTC samples to the interaction of SYBR Green I with focused ssDNA, which yields a non-negligible fluorescence[32].

To demonstrate the use of on-demand encapsulation for downstream reactions, we encapsulated the leading band (Fig. 4c), ejected the droplet from the encapsulation region (Fig. 4d) and manually extracted the $240~pL$ analyte droplet from the reservoir of the encapsulation device using a pipette. The encapsulation was not affected by the asymmetry in the separation channel (1.9 cm to the left of the encapsulation region, and 4.2 mm to the right), suggesting that this method can be used irrespective of the upstream separation geometry. After encapsulation, we performed parallel PCR detection on a single droplet by mixing it with a volume of PCR master mix sufficient for nine individual reactions. We then split this volume into nine vials and added β-actin primers, BRAF15 primers or no primers to three vials each (*i.e.,* triplicate experiments).

Figure 4e shows the PCR amplification results for three separate droplets, with nine reactions for each droplet (triplicates of each of the three primer conditions). The samples with β-actin primers amplify after just 11 cycles, confirming that the amplicon is present at high concentration in the leading band. However, vials containing BRAF15 primers did not show amplification even after 35 cycles, indicating that genomic DNA is successfully separated from the amplicon, as it does not focus in the leading band. Although Fig. 4b shows that primers are expected to focus in the leading band, we observed that this band did not amplify in samples to which no primers were added. This indicates that all the primers are converted to amplicon in the first round of PCR, so that none remain for subsequent amplification in the second round.

The specificity of the amplification is corroborated by the melt curve, whose derivative is shown in Figure 4f. The melt curves for each family (amplification pre-ITP or post-encapsulation, shown as dashed and solid lines, respectively) are normalized to the maximum of the positive β-actin sample in that family. This maximum occurs around 85.5 °C for both the pre-ITP and post-encapsulation β-actin amplification. The measured melting temperature is in good agreement with the $81 - 85$ °C range calculated using the POLAND tool[33]. By contrast, no peak is visible at 79 °C, the melting temperature of BRAF15 products.

## 6. SUMMARY AND CONCLUSIONS

We presented a method for retention and manipulation of electrokinetically focused and separated analytes using a two-phase system. This method uses capillary pinning in hydrophobic secondary channels to hold an oil interface during electrokinetic separation of analytes in the primary channel. As the desired analyte band passes between the secondary channels, a brief pressure pulse is applied to the oil/water interface, bursting the capillary valves and encapsulating the analyte in a droplet.

To illustrate the use of this method we developed a three-channel geometry for the encapsulation of single analyte bands. We discussed geometrical parameters that are central to the functioning of this device, with an emphasis on those related to the displacement of the aqueous phase during oil injection. A low displaced volume is desirable as it enables encapsulation with minimal dispersion of the analyte band. This can be achieved by a geometrically defined transition of the oil interface between two stable pinning positions, which is determined by six geometrical parameters. We showed that the geometry is highly adaptable and can be extended to encapsulate multiple analytes with the addition of opposing channels.

We used the three-channel geometry to encapsulate DNA oligonucleotides after isotachophoresis focusing. After encapsulation, we held a 240 pL aqueous droplet containing the focused DNA in place and showed that the concentration rapidly decays to around 60% of the initial maximum before remaining constant for tens of minutes. After 20 minutes, the measured concentration was 22× higher than the concentration expected from the equation for 1-D diffusion without encapsulation. A further increase in the retained concentration may be achieved by reducing the length of the encapsulation zone, although automated feedback control of the pressure lines is recommended in this case.

In contrast with valve-based trapping methods, the encapsulation approach allows the analyte droplet to be manipulated for *e.g.* off-chip analysis. We demonstrated this by encapsulating ITP-purified amplicon focused from a PCR reaction mixture, and subsequently performing nine parallel off-chip detection reactions from the single droplet. These results showed that the encapsulated droplet contained a high concentration of amplicon and no contaminating genomic DNA, enabling downstream reactions that require a highly purified amplicon fraction.

Although our demonstration focused on an off-chip assay, on-demand encapsulation may also be coupled to on-chip assays in a droplet format. The use of droplet-based assays would enable a range of reactions that are impossible or inefficient under the constraints given by electrokinetic focusing or separation. Examples include multi-step immunoassays with a different optimal pH for the focusing and reaction steps[12,34] or single-step assays that use complex mixtures of reagents with a wide range of positive and negative electrophoretic mobilities (*e.g.* loop-mediated isothermal amplification[35] or PCR).

The key technical challenge in combining encapsulation with other droplet-based methods lies in developing surfaces with wetting properties that enable pinning at the capillary valves while also enabling controlled formation of reagent droplets. This may, for instance, be achieved by applying different surface treatments to different parts of the same encapsulation device, or by transferring the encapsulated droplet to a separate chip dedicated to droplet manipulation.

We also note that a high aqueous contact angle is not always desired. For example, encapsulated analyte zones that do wet the channel walls could be delivered to surface-based assays and sensors (*e.g.* exposed electrochemical sensors[36] or surface-enhanced Raman spectroscopy[37]) that may be sensitive to external electric fields and are thus located outside the separation channel. The ability to tune the contact angle of the analyte zone after on-demand encapsulation may also provide a way to deliver a plug containing concentrated analytes to sensing surfaces, and to remove it for subsequent read-out.

While we demonstrated the encapsulation method using ITP separation and focusing, we expect the method to be relevant to other electrokinetic methods such as CZE and IEF, as it is largely independent of the composition of the aqueous separation phase (except for those containing high concentrations of surfactants, for which optimization will be needed). The operation of the device is also largely independent of the upstream or downstream geometry of the separation channel, making the encapsulation method applicable to a range of separation geometries. We believe that this versatility makes on-demand encapsulation an appealing new way to bridge the gap between upstream electrokinetic separation techniques and downstream detection or analysis, and that it will enable a range of assays that until now were incompatible with electrokinetics.

## 7. ACKNOWLEDGEMENTS
This work was supported by the Initial Training Network, Virtual Vials, funded by the FP7 Marie Curie Actions of the European Commission (FP7-PEOPLE-2013-ITN-607322). We thank Dr. Onur Gökçe and Federico Paratore for valuable discussions and technical help. Dr. Emmanuel Delamarche and Dr. Walter Riess are acknowledged for their continuous support.